\documentstyle[psfig,prl,aps,multicol]{revtex}
\begin{document}
\title{Charged magnetoexcitons in two-dimensions: \\
         Magnetic translations and families of dark states}
\author{A. B. Dzyubenko\cite{ABD}}
\address{
 Institut f\"{u}r Theoretische Physik, J.W. Goethe-Universit\"{a}t,
 60054 Frankfurt,  Germany }
\author{A. Yu. Sivachenko}
\address{The Weizmann Institute of Science, Rehovot 76100, Israel
}
\date{\today}
\maketitle
\begin{abstract}
We show that optical transitions of charged excitons
in semiconductor heterostructures
are governed in magnetic fields by a novel exact selection rule,
a manifestation of magnetic translations.
It is shown that the spin-triplet ground state of the quasi-two-dimensional
charged exciton $X^-$ --- a bound state of two electrons and one hole ---
is optically inactive in photoluminescence at finite magnetic fields.
Internal bound-to-bound $X^-$ triplet transition has a specific
spectral position, below the electron cyclotron resonance, and is strictly
prohibited in a translationally-invariant system.
These results allow one to discriminate between localized and free
charged excitons.
\end{abstract}


%\narrowtext
%\twocolumn

\vspace*{-1cm}

\parskip0pt

\begin{multicols}{2}

In quasi-two-dimensional (quasi-2D) electron-hole \mbox{($e$--$h$)}
systems with low density of particles, a variety of hydrogenic few-particle
complexes can be formed.
Optical spectroscopy in magnetic fields $B$ is one of the basic tools for
studying such complexes.
Recently, much experimental \cite{Fink,Shlds,Fink2,Yak,Kuk,Loc,Hayne}
and theoretical \cite{Hawr,AHM,Whit97,Stebe,Quinn}
attention has been devoted to studying negatively $X^-$ ($2e$--$h$)
and positively $X^+$ ($2h$--$e$) charged excitons in magnetic fields $B$.
These complexes are often considered to be semiconductor analogs of the
hydrogen atomic $H^-$ and molecular $H_2^+$ ions, respectively.
In $B$, in addition to the spin-singlet, higher-lying spin-triplet
bound states of $X^-$ and $X^+$ develop \cite{Fink,Shlds}.
The question as to whether these complexes are mobile and free to move,
or are localized --- by single donor impurities \cite{Kuk,Hayne,Whit97},
disorder due to long-range fluctuating potential of remote
donors \cite{Loc} etc. --- is a matter of current controversy.
To explore these issues, we theoretically address
from first principles the following question:
{\em Are there fundamental differences in optical transitions
between mobile and localized charged $e$--$h$ systems in magnetic fields?}

For a one-component translationally-invariant interacting electron system
in $B$, the well-known Kohn theorem \cite{Kohn} states that
intraband transitions can occur only at the bare electron cyclotron resonance
($e$--CR) energy $\hbar\omega_{\rm ce}= \hbar eB/m_e c$.
This is a consequence of the center-of-mass (CM) separation from internal
degrees of freedom in $B$. For $e$--$h$ systems such separation is not
possible and the CM and internal motions are coupled in $B$ \cite{G&D,Simon}.
However, for any system of charged particles in a uniform $B$
an exact symmetry --- magnetic translations --- exists
(\cite{Simon} and references therein). It has been used to study
the motion of atoms and ions in constant magnetic and electric fields
\cite{Hirsch}. This symmetry, however,
has not been identified in previous theoretical work on charged excitons
in $B$. In this work, we introduce for charged semiconductor $e$--$h$
complexes in $B$ an exact classification of states, which is based on
magnetic translations \cite{Simon}.
In this scheme, in addition to the total orbital
angular momentum projection $M_z$ and spin of electrons $S_e$ and holes $S_h$,
an exact quantum number, the oscillator quantum number $k$, appears.
Surprisingly, only very general consideration of radiation processes in $B$
using magnetic translations has been given \cite{Simon}.
To our knowledge, no selection rule associated with $k$ has been established
for dipole-allowed magneto-optical transitions.
We show here that $k$ is strictly conserved in the
intraband and in interband magneto-optical transitions.
This leads to striking spectroscopic consequences for charged
excitons.

  Consider a many-body Hamiltonian of interacting particles of
charges $e_i$ in a magnetic field ${\bf B}=(0,0,B)$
\begin{equation}
                \label{H} %(1)
   H = \sum_i \frac{ \hat{ \bbox{ \pi }}_i^2}{2m_i}
          + \case{1}{2} \sum_{i \ne j} U_{i j}({\bf r}_i-{\bf r}_j) \, ,
\end{equation}
here
$\hat{\bbox{\pi}}_i = -i\hbar \bbox{\nabla }_i -
\frac{e_i}{c} {\bf A}({\bf r}_i)$
and potentials of interparticle interactions $U_{ij}$
can be rather arbitrary.
In the symmetric gauge ${\bf A} = \frac12 {\bf B} \times {\bf r}$
the total angular momentum projection $M_z$, an eigenvalue
of $\hat{L}_z=\sum_i ({\bf r}_i \times -i\hbar\bbox{\nabla }_i)_z$,
is an exact quantum number. In a uniform ${\bf B}$
the Hamiltonian (\ref{H}) is also invariant under
a group of magnetic translations whose generators are the components
of the operator $\hat{\bf K} = \sum_{i} \hat{\bf K}_i$,
where $\hat{\bf K}_i =
\hat{\bbox{ \pi }}_i - \frac{e_i}{c} {\bf r}_i \times {\bf B}$
and  $[\hat{K}_{ip}, \hat{\pi}_{iq}]= 0$, $p,q=x,y$ \cite{G&D,Simon,Hirsch}.
$\hat{\bf K}$ is an exact integral of the motion $[H, \hat{\bf K}]=0$.
Its components commute as
\begin{equation}
        \label{comK}
 [\hat{K}_x, \hat{K}_y] = - i \frac{\hbar B}{c} Q \quad , \quad
 \quad Q \equiv \sum_i e_i  \, .
\end{equation}
For neutral complexes (excitons, biexcitons) $Q=0$
and classification of states in ${\bf B}$ is due to the two-component
continuous vector --- the 2D magnetic momentum
${\bf K}= (K_x,K_y)$ \cite{G&D,Simon}.
For charged systems $Q  \ne 0 $ and the components
of $\hat{\bf K}$ do not commute. This determines the macroscopic
Landau degeneracy of eigenstates of (\ref{H}).
Using a dimensionless operator
$\hat{{\bf k}} = \sqrt{c/\hbar B |Q|} \, \hat{\bf K}$
whose components are canonically conjugate,
one obtains raising and lowering Bose ladder operators
$\hat{k}_{\pm}= (\hat{k}_x  \pm i \hat{k}_y)/\sqrt{2}$
such that $[\hat{k}_{-}, \hat{k}_{+}]=Q/|Q|$.
Therefore,
$\hat{{\bf k}}^2 = \hat{k}_{+} \hat{k}_{-} + \hat{k}_{-} \hat{k}_{+}$
has the oscillator eigenvalues $2k+1$, $k=0, 1, \ldots$.
Since $[\hat{{\bf k}}^2, H]=0$ and $[\hat{{\bf k}}^2, \hat{L}_z]=0$,
the exact charged eigenstates of (\ref{H}) can be simultaneously labeled
by the discrete quantum numbers $k$ and $M_z$ \cite{Simon}.
For charged $e$--$h$ complexes in $B$ the labelling therefore is
$|k M_z S_e S_h \nu \rangle$,
where $\nu$ is the ``principal'' quantum number, which
can be discrete (bound states) or continuous (unbound states forming
a continuum); concrete examples are given below.
The $k=0$ states are {\em Parent States} (PS's) within a degenerate manifold.
All other daughter states $k=1,2,\ldots$
in each $\nu$-th family can be generated out of the PS: for, e.g., $Q<0$
\begin{equation}
        \label{DS}
  |k,M_z-k,S_e S_h \nu \rangle =
  \frac{1}{\sqrt{k!}} \hat{k}_{-}^k |0, M_z, S_e S_h \nu \rangle \, ,
\end{equation}
where we have used $[\hat{L}_z,\hat{k}_{\pm}]= \pm \hat{k}_{\pm}$.
The values of $M_z$ that the PS's have are determined by
particulars of interactions and cannot be established {\it a priori}
(cf.\ with 2D electron systems in strong $B$ \cite{Kivelson}).

In the dipole approximation the photon momentum is negligibly small.
Therefore, the quantum number $k$ should be conserved in
intra- and inter-band magneto-optical transitions.
Let us establish this selection rule formally.
For internal intraband transitions in the Faraday geometry
(light propagating along ${\bf B}$) the Hamiltonian of the interaction
with the radiation of polarization $\sigma^{\pm}$ is of the form
$\hat{V}^{\pm} = \sum_{i}
(e_i {\cal F}_0 \hat{ \pi }_{i\pm}/ m_i \omega) e^{-i \omega t}$,
where ${\cal F}_0$ is the radiation electric field,
$\hat{\pi}_{i\pm} = \hat{\pi}_{ix} \pm i \hat{\pi}_{iy}$
(e.g., \cite{Kohn}).
Conservation of $k$ follows from the commutativity
$[\hat{V}^{\pm}, \hat{\bf K}]=0$ \cite{rel}.
(In fact the perturbation $\hat{V}=F(\hat{\bbox{ \pi }}_i,t)$
can be an arbitrary function of kinematic momentum operators
$\hat{\bbox{ \pi }}_i$ and time $t$, corresponding, e.g., to other
geometries and polarization.)
Other usual selection rules are conservation of spins
$S_e$, $S_h$ and $\Delta M_z= \pm 1$ for the envelope function in the
$\sigma^{\pm}$ polarization. This means that the PS's should be
connected by the dipole transition, i.e., have
proper spins and $M_{z}' - M_{z} = \pm 1$.
Indeed, for the transition dipole matrix element between the daughter
states in the $k'$-th and $k$-th generations we have
\begin{eqnarray}
        \label{D}
{\cal D}_{\nu' \nu} & = &
\langle k',M_z'-k',S_e S_h \nu'| \hat{V}^{\pm} |k,M_z-k,S_eS_h \nu \rangle \\
& = & \nonumber
\delta_{k',k} \delta_{M_z',M_z \pm 1}
\langle 0,M_z',S_eS_h\nu'| \hat{V}^{\pm} |0,M_z,S_eS_h\nu \rangle \, .
\end{eqnarray}
Here we have used (\ref{DS}) and the operator algebra
$[\hat{V}^{\pm} , \hat{k}_{-} ] = [\hat{V}^{\pm} , \hat{k}_{+} ] =0$,
$[\hat{k}_{+},\hat{k}_{-} ]=1$.
From (\ref{D}) it is clear that ${\cal D}_{\nu' \nu}$ is the same
in all generations and, thus, characterizes the two families of states.
Similar considerations apply to interband transitions with
$e$--$h$ pair creation or annihilation:
The interaction with the radiation field is described by the luminescence
operator
$\hat{\cal L}_{\rm PL}= p_{\rm cv} \int \! d{\bf r} \,
\hat{\Psi}^{\dagger}_{e}({\bf r}) \hat{\Psi}^{\dagger}_{h}({\bf r})
+ \mbox{H.c.}$, where $p_{\rm cv}$ is the interband momentum
matrix element (e.g., \cite{Haug}).
Here we have the commutativity $[\hat{\cal L}_{\rm PL},\hat{\bf K}]=0$,
so that $k$ is conserved. Due to the change of the Bloch parts in this case,
the usual selection rule $\Delta M_z= 0$ holds for the envelope functions.

Conservation of $k$ constitutes an exact selection rule
for the dipole-allowed magneto-optical transitions.
It is applicable to any charged $e$--$h$ system in $B$.
Mobile charged excitons $X^-$, $X^+$,
charged multiple-excitons $X^{-}_{N}$ \cite{AHM,Quinn},
are particular examples.
In some limiting cases $k$ can be directly related to the center of
the cyclotron motion of the complex as a whole \cite{Simon,Hirsch}.
This gives some physical insight into its conservation.
In the derivation above we only used translational invariance
in the plane perpendicular to ${\bf B}$. Therefore, conservation of $k$
holds in arbitrary magnetic fields and for systems of different
dimensionality. Importantly, it is also applicable to semiconductors
with {\em complex} valence band. Indeed, $k$ is a good quantum number
for the Luttinger Hamiltonian, while $M_z$ is replaced by
the combination ${\cal M}_z= M_z+S_{\rm hz}+S_{\rm ez}$
involving the $e$- and $h$- spin projections;
${\cal M}_z$ is conserved in the usual axial approximation
(e.g., Ref.\ \cite{Haug}, p.\ 48).

    To make further discussion more concrete, we consider the
strictly-2D $e$--$h$ system in the limit of high $B$ \cite{AHM,Whit97},
when $\hbar\omega_{\rm ce}, \hbar\omega_{\rm ch} \gg
E_0 = \sqrt{\pi/2} \, e^2/\epsilon l_B$
and mixing between Landau levels (LL's) can be neglected;
$l_B = (\hbar c/e B)^{1/2}$. $E_0$ is the characteristic energy of Coulomb
interactions, the only energy scale in the problem.
The basis for the $X^-$ states \cite{rmrk} in the electron and hole LL's
$(N_eN_h)$ is of the form
$\phi^{(e)}_{n_1 m_1}({\bf r}_e) \,
 \phi^{(e)}_{n_2 m_2}({\bf R}_e) \,
 \phi^{(h)}_{N_{h}m_{h}}({\bf r}_{h})$
and includes different three-particle $2e$--$h$ states
such that the total angular momentum projection
$M_z= n_1 + n_2 -m_1 -m_2 - N_h + m_h$,
and LL's $N_e=n_1+n_2$, $N_h$ are fixed \cite{Dz_PLA,Dz_PLA93}.
Here $\phi^{(e,h)}_{n m}$ are the $e$- and $h$- single-particle
factored wave functions in $B$ (e.g., \cite{Simon,Hirsch});
$n$ is the LL quantum number and $m$ is the single-particle
oscillator quantum number ($m_{\rm ze}=-m_{\rm zh}= n-m$).
We use the electron relative
${\bf r}_e = ({\bf r}_{e1} - {\bf r}_{e2})/\sqrt{2}$
and CM
${\bf R}_e = ({\bf r}_{e1} + {\bf r}_{e2})/\sqrt{2}$  coordinates.
The electron relative motion angular momentum $n_1-m_1$ should be even (odd)
in the electron spin-singlet $S_e=0$ (triplet $S_e=1$) state.
To make this basis compatible with magnetic translations, i.e., to fix
$k$, an additional canonical transformation diagonalizing
$\hat{{\bf k}}^2$ should be performed; details will be given elsewhere.
The method of analytical calculation of the Coulomb matrix elements has
been described in \cite{Dz_PLA,Dz_PLA93}.

The calculated three-particle $2e$--$h$ eigenspectra with electrons in the
triplet $S_e=1$ state in two lowest LL's are shown in Fig.\,1.
The spectral properties of the charged three-body problem in strong $B$
is interesting in itself \cite{Simon,Hirsch}.
Generally, the eigenspectra associated with each LL consist of bands of
finite widths $\sim E_0$. The states within each such band form a
continuum corresponding to the extended motion of a {\em neutral}
magnetoexciton (MX) as a whole with the second electron in a scattering state
(on average at infinity from the MX).
For example, the continuum in the ($N_eN_h$)=(10) LL consists of the MX
band of width $E_0$ extending down in energy from 
\parbox{\columnwidth}{
\psfig{file=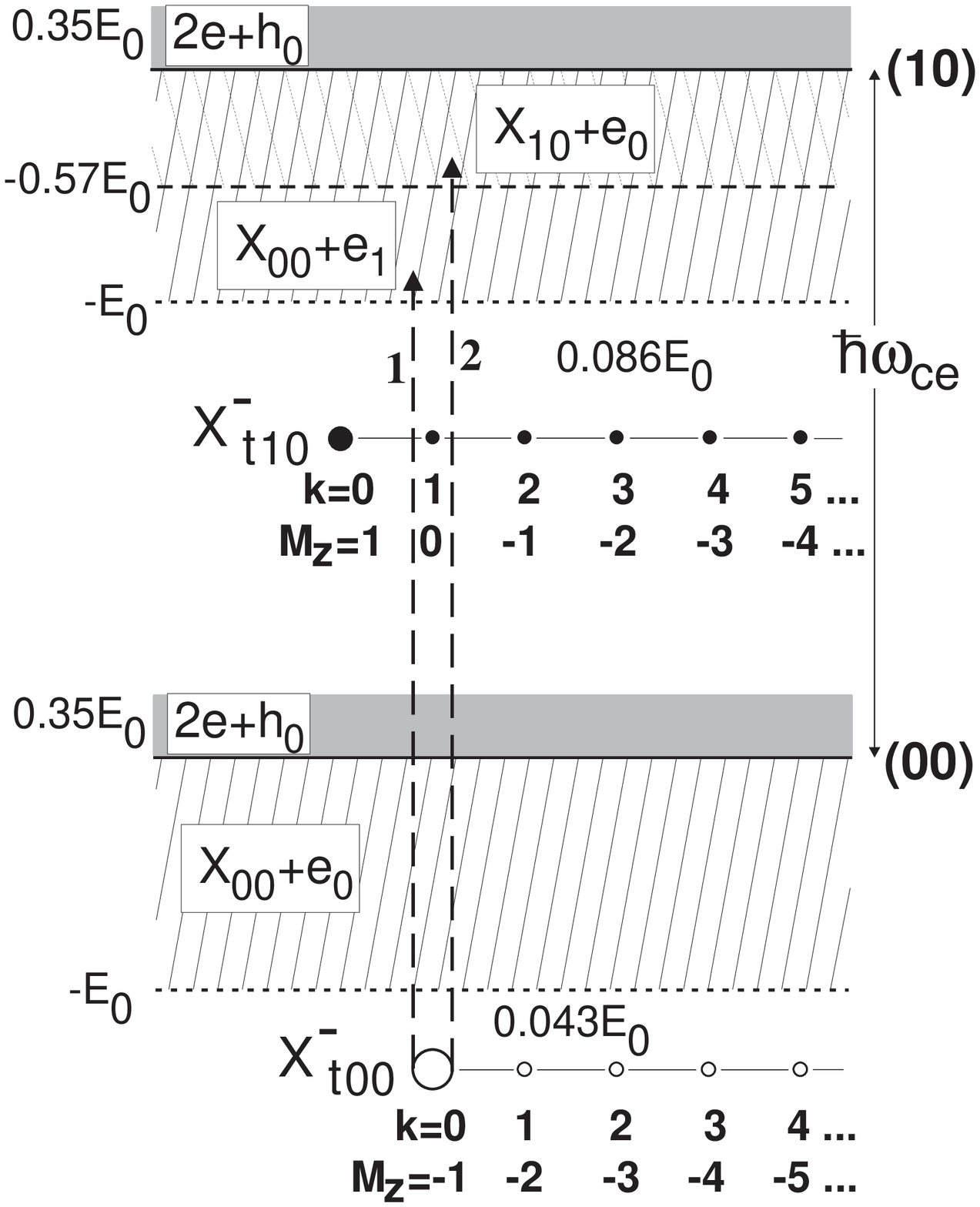,width=\columnwidth,clip=}

%\vskip1mm

\small { FIG. 1.}
Schematic drawing of bound and scattering electron triplet $2e$--$h$ states
in the lowest LL's ($N_eN_h$)=(00), (10).
Large (small) dots correspond to the bound parent $k=0$
(daughter $k=1,2,\ldots$) $X^-$ states.
Allowed internal transitions must satisfy $\Delta M_z = 1$ and $\Delta k = 0$.

\vskip3mm

}
the free $(10)$ LL. This corresponds to the $1s$ MX in zero 
LL's \cite{L&L80}
plus a scattered electron in the first LL, labeled $X_{00}+e_1$.
(A similar continuum exists in zero LL's.)
In addition, there is another MX band of width $0.574 E_0$ also
extending down in energy from the free (10) LL. This
corresponds to the $2p^+$ MX \cite{L&L80}
plus a scattered electron in the zero LL, labeled $X_{10}+e_0$.
Moreover, there is a band above each free LL
(labeled $2e+h_0$ in Fig.\,1) originating from the bound
internal motion of two 2D electrons in $B$ (cf.\ \cite{Kivelson}).
Bound $X^-$ states lie outside the continua.
In the strictly-2D high-$B$ limit the only family of
bound $X^-$ states in zero LL's
is the triplet $X^-_{t00}$. There are no bound
singlet $X^-_s$ states \cite{AHM,Whit97} in contrast to the $B = 0$ case.
The obtained $X^-_{t00}$ binding energy $0.043 E_0$ is in agreement
with \cite{AHM,Whit97}. In the next electron LL
there are no bound singlet $X^-_s$ states, and only one family of
bound triplet states $X^-_{t10}$.
The $X^-_{t10}$ binding energy is $0.086 E_0$, twice that of the $X^-_{t00}$.
This is due to the fact that the two electrons in the triplet $X^-_{t10}$
state can occupy the single-particle states with zero $e$--$h$ relative
angular momenta $1s$ (zero LL) and $2s$ (first LL).
This enhances the $e$--$h$ attraction relative to the ground
$X^-_{t00}$ state in which electrons can occupy
an antisymmetric combination of 
the $1s$ and $2p^-$ single-particle states in zero LL.
\parbox{\columnwidth}{
\psfig{file=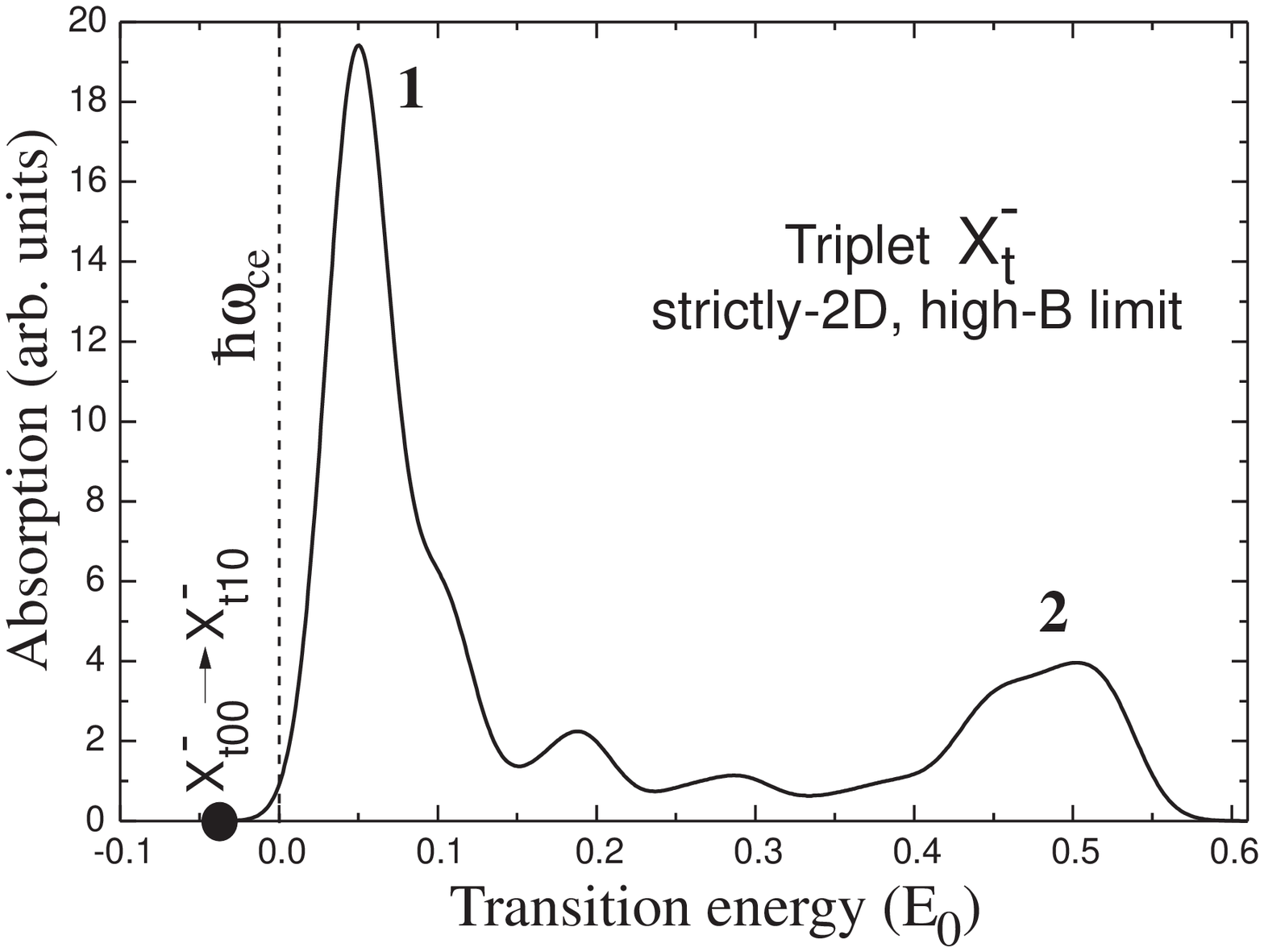,width=\columnwidth,clip=}

\vskip1mm

\small {FIG. 2.}
 Energies
(in units of $E_0= \protect\sqrt{\pi/2} \, e^2/\epsilon l_B$,
counted from $\hbar\omega_{\rm ce}$)
and dipole matrix elements of the internal transitions
corresponding to Fig.\,1.
The filled dot shows the position of the forbidden
$X^-_{t00} \protect\rightarrow X^-_{t10}$ transition.
Spectra have been convoluted with the Gaussian of the $0.02E_0$ width.

\vskip2mm

}

 We first discuss internal $X^-$ triplet transitions.
In the $\sigma^+$ polarization the inter-LL $\Delta N_e = 1$
transitions are strong and gain strength with $B$:
Transitions with $\Delta N_e \neq 1$ are only due to LL mixing and
weak as $[E_0/\hbar\omega_{\rm ce(h)}]^2 \sim B^{-1}$.
Both bound-to-bound $X^-_{t00} \rightarrow X^-_{t10}$
and photoionizing bound-to-continuum transitions are possible.
For the latter, due to the rich structure of the continuum,
two exact selection rules (\ref{D})
are easily simultaneously satisfied.
As a result, the photoionizing absorption spectra
have intrinsic linewidth $\sim E_0$ with two prominent peaks
above the $e$--CR (Fig.\,2).
These peaks are associated with high densities of states
at the edges of the two MX bands indicated in Fig.\,1.
Transitions to the $2e+h_0$ band have extremely
small oscillator strengths.
Most of these qualitative features of photoionizing transitions
are preserved at finite fields and confinement, where both the
triplet and singlet bound $X^-$ states exist.
This has been shown by high-accuracy calculations for realistic
GaAs/GaAlAs quantum wells at $B > 8$\,T, which are
confirmed in recent experiments and will be reported elsewhere \cite{tobe}.
Here we are interested in the bound-to-bound
$X^-_{t00} \rightarrow X^-_{t10}$ transition.
Note first that since the final state is more strongly bound,
this transition has a specific spectral position ---
it lies below the $e$--CR energy $\hbar\omega_{\rm ce}$.
However, in a translationally invariant system it is
{\em strictly prohibited\/}.
Indeed, the $X_{t00}$ PS (with $k=0$) has $M_z=-1$, while
the $X_{t10}$ PS has $M_z'=1$ (Fig.\,1).
It follows then that the selection rules (\ref{D})
$\Delta k=0$ and $\Delta M_z =1$
cannot be simultaneously satisfied.
This also holds at finite $B>8$\,T and in quasi-2D quantum wells \cite{tobe}.
Localization of charged excitons breaks translational invariance and
relaxes selection rules. As a result, the bound-to-bound
$X^-_{t00} \protect\rightarrow X^-_{t10}$ transition
develops {\em below} the $e$--CR. Such a peak is a tell-tale mark of
localization of charged triplet excitons.
The strong triplet $T^-$ transition of the $D^-$ center
(two electrons bound by a donor ion), which was predicted
theoretically \cite{Dz_PLA} and observed experimentally
\cite{Int_D}, can be thought of as one of the possible limiting cases,
namely, when the hole is completely localized.

Consider now photoluminescence (PL) from the triplet ground state
$X^-_{t00} \rightarrow \mbox{\rm photon} + e^-_{n}$
with the electron in the $n$-th LL in the final state;
$n=1,2,\ldots$ correspond to shake-up processes in the PL
\cite{Fink2,Yak,Kuk}.
The PL selection rules are $\Delta k=0$ and  $\Delta M_z=0$.
The triplet $X^-_{t00}$ ground PS with $k=0$ has $M_z=-1$
(also at finite $B>8$\,T and in quasi-2D quantum wells \cite{Whit97,tobe}),
while the electron in the $n$-th LL, with the factored wave function
$\phi^{(e)}_{n m}$, has $m_z=n-m$.
The corresponding optical matrix element
for transition to any LL $n \geq 0 $ is zero:
$\langle \phi^{(e)}_{n m}| \hat{\cal L}_{\rm PL} |X^-_{t00(M_z=-1,k=0)}\rangle
\sim \delta_{m,k=0} \delta_{n-m,-1}$.
This means that the ground triplet state of an isolated $X^-_t$
is optically inactive --- {\em dark in PL}.
In the strictly-2D high-$B$ limit this also follows \cite{AHM}
from the ``hidden symmetry'' in $e$--$h$ systems \cite{hidden}.
Our result is much more general.
Indeed, as discussed above, quasi-2D effects, $e$--$h$ asymmetry,
admixture of higher LL's, and the complex character of the valence band
break neither axial nor translational symmetry.
Therefore, even in the presence of these effects, the triplet
stays dark --- as long as the ground $X^-_t$ PS has $M_z<0$.
This exact result was overlooked in \cite{AHM,Whit97};
very small but finite $X^-_t$ oscillator strengths obtained
in \cite{AHM,Whit97} are in fact artifacts coming from finite-size
calculations.
Note that the quasi-2D $X^-_s$ singlet ground PS has $M_z=0$
\cite{Whit97,Stebe,tobe} and is optically active in PL:
$\langle \phi^{(e)}_{n m}| \hat{\cal L}_{\rm PL} |X^-_{s00(M_z=0,k=0)} \rangle
\sim \delta_{m,0} \delta_{n,0}$.
We see, however, that the shake-up processes are prohibited in
PL from the isolated singlet ground state $X^-_s$.
The question now remains why in fact the $X^-_t$ ground state
is visible in experiment in $B$ \cite{Fink,Shlds,Fink2,Yak,Kuk,Loc,Hayne}
and the singlet $X^-_s$ shake-up processes are commonly
observed in PL --- even at very low
densities of excess free carriers \cite{Fink2,Kuk}?
Our results show that there should be mechanisms breaking the
underlying exact translational and rotational symmetries.
We interpret this as an indication toward localization of charged
excitons in $B$.
More theoretical and experimental work is needed here to establish,
in particular, the regime of localization of charged excitons.

In conclusion, we have shown that due to magnetic translations,
dipole-allowed transitions of charged mobile semiconductor complexes
are governed in magnetic fields $B$ by a novel
exact selection rule.
Some experimentally observed features in interband
photoluminescence of quasi-2D charged excitons $X^-$ in $B$ cannot be
explained without accounting for symmetry-breaking effects,
an indication toward localization.
The appearance of the peak below the electron cyclotron
resonance, corresponding to the internal bound-to-bound $X^-$ triplet
transition, is a characteristic mark associated with breaking of
translational invariance. We propose using this as a tool for studying
the extent of $X^-$ localization.

We thank I.~Bar-Joseph, B.D.~McCombe,
D.M.~Whittaker, and D.R.~Yakovlev for useful discussions.
ABD is grateful to the Humboldt Foundation for research support.

\end{multicols}


\begin{references}
\bibitem{rmrk} In the high-$B$ limit, the $X^{+}$
eigenspectra are obtained from those of $X^{-}$
with the substitution $M_z \protect\rightarrow - M_z$.
\end{references}
\end{document}